\documentclass[a4,12pt]{article}

\usepackage{amsmath,amssymb,amsfonts,graphics,graphicx,amscd,amsfonts,epsf,epsfig,color}
\makeatletter
\@addtoreset{equation}{section}
\renewcommand{\theequation}{\thesection.\@arabic\c@equation}
\makeatother

\makeatletter
\renewcommand\appendix{\par
  \setcounter{section}{0}%
  \setcounter{subsection}{0}%
  \gdef\thesection{Appendix \@Alph\c@section }
  \renewcommand{\theequation}
  {\Alph{section}.\arabic{equation}}
}
\makeatother

\newcommand{\ba}{\begin{eqnarray}}
\newcommand{\ea}{\end{eqnarray}}

\newcommand{\cO}{{\cal O}}
\newcommand{\cP}{{\cal P}}
\newcommand{\cQ}{{\cal Q}}
\newcommand{\nt}{\notag\\}

\newcommand{\vect}[2]{\vec{#1}^{\;\!#2}}

\date{}
\begin{document}

\begin{titlepage}

\begin{flushright}
MISC-2015-01
\end{flushright}

\begin{center}
  {\LARGE
 Thermodynamics of Intersecting Black Branes 
 \\[+5pt] from Interacting Elementary Branes
  }
\end{center}
\vspace{0.2cm}
\baselineskip 18pt 
\renewcommand{\thefootnote}{\fnsymbol{footnote}}

\begin{center}
Takeshi {\sc Morita}${}^{a}$\footnote{%
E-mail address: morita.takeshi@shizuoka.ac.jp
}, 
Shotaro S{\sc hiba}${}^{b}$\footnote{%
E-mail address: sshiba@cc.kyoto-su.ac.jp}

\renewcommand{\thefootnote}{\arabic{footnote}}
\setcounter{footnote}{0}

\vspace{0.4cm}

{\small\it

${}^a$ Department of Physics,
Shizuoka University, \\
836 Ohya, Suruga-ku, Shizuoka 422-8529, Japan

${}^b$ Maskawa Institute for Science and Culture, Kyoto Sangyo University, \\
Kamigamo-Motoyama, Kita-ku, Kyoto 603-8555, Japan.

}

\end{center}


\vspace{1.5cm}

\begin{abstract}

If an Einstein-Maxwell-Dilaton system admits the extreme brane solution in which no force works between the parallel branes, the collective motion of nearly parallel branes exhibits the thermodynamical properties which are coincident with those of the corresponding black branes at low energy regime  (up to unfixed numerical factors). 
Hence it may provide the microscopic description of the black branes ($p$-soup proposal).
This fact motivates us to test this proposal in the intersecting black branes which have multiple brane charges and/or momentum along the brane direction.
We consider the case that the multiple branes satisfy the intersection rule and feel no force when they are static, and find the agreement to the black hole thermodynamics.
\end{abstract}


\end{titlepage}

\newpage
\baselineskip 18pt
\section{Introduction}

Understanding the microscopic origin of black hole thermodynamics is one of the most important outstanding problem in theoretical physics.
Recently Ref.\cite{Morita:2013wfa}  proposed the following microscopic description of the near extremal black branes in the Einstein-Maxwell-Dilaton system. 
This system admits the extreme brane solution in which no force works between the parallel branes if we tune the dilaton coupling suitably.
In case these branes are nearly parallel or slowly moving, they start interacting with each other and
the low energy effective action for these branes is given as
\begin{align}
S=\int dt \left(
\sum_{i=1}^{N}  \frac{m}{2} \vect{v}{2}_i+
\sum_{i\neq j}^{N}  \frac{\kappa^2_{4} m^2}{16\pi} \frac{\vect{v}{4}_{ij}}{\vec{r}_{ij}}
+\cdots\right).
\label{single-brane}
\end{align}
Here we consider 0-branes in four dimensions as a simple example. $N$ is the number of the branes, $\vec{v}_i$ is velocity of the $i$-th brane, $m$ is mass of the brane and $\kappa_{4}$ is the gravitational coupling.
Also $\vec{r}_{ij}\equiv \vec{r}_i-\vec{r}_j$ and $\vec{v}_{ij}\equiv \vec{v}_i-\vec{v}_j$ denote the relative position and the relative velocity of the $i$-th and $j$-th branes.
The first term is the ordinary non-relativistic kinetic term and the second term is the interactions between the branes which vanish when the branes are relatively static.

If we assume that the branes compose a bound state due to the interactions, we can estimate the thermodynamical properties of the bound state by applying the virial theorem to the effective action (\ref{single-brane}).
In \cite{Morita:2013wfa}, we found that the branes are strongly coupled in the bound state and the  thermodynamical properties are coincident with the corresponding black brane thermodynamics in the near extremal regime (up to unfixed numerical factors).
It suggests that the black hole thermodynamics may be explained via the effective action (\ref{single-brane}) microscopically.
We call this conjecture ``$p$-soup proposal".

This proposal works in the various BPS branes in superstring and M-theory, and we can explain the black hole thermodynamics of the black brane solutions through the effective actions for these branes similar to (\ref{single-brane})  \cite{Wiseman:2013cda, Morita:2013wla, Morita:2014ypa}.
(The related studies have been done in  \cite{Horowitz:1997fr,Li:1997iz,Banks:1997tn,Li:1998ci,Smilga:2008bt}.)
Moreover, this proposal works even in the D1-D5(-P) system \cite{Morita:2014cfa}.
Since this system \cite{Callan:1996dv,David:2002wn}  involves the different species of the branes and the supersymmetry is less than the single brane case, it is non-trivial that the $p$-soup proposal works in such a system.

This motivates us to investigate general multiple species of branes in the Einstein-Maxwell-Dilaton system.
We set these multiple species of branes so that the forces between the branes vanish when they are static. 
It happens when the branes satisfy so-called ``intersecting rule'' \cite{Argurio:1997gt}, and the D1-D5(-P) system also satisfies it.
Then we derive the effective action for these branes similar to (\ref{single-brane}), and, by analyzing this action, we find that the bound state of these branes exhibits thermodynamical properties corresponding to the black hole thermodynamics.
Thus the $p$-soup proposal works in this system, and we believe that it captures the fundamental natures of the black hole microstate.

This paper is organized as follows.
In Sec.\,\ref{sec:setup} we review Einstein-Maxwell-Dilaton theory and clarify our setup of intersecting brane systems.
In Sec.\,\ref{sec:twobrane} we discuss intersecting brane systems of only two species of branes. This simple example would be helpful to understand our discussion without complicating the story.
The general intersecting brane systems are discussed in Sec.\,\ref{sec:general}, but the details of calculation are shown in \ref{app:eff}.
Sec.\,\ref{sec-conclusion} is the discussions including possible applications to string theory.
The corresponding black branes in supergravity and their thermodynamics are discussed in \ref{app:thermo}.

\section{Intersecting brane system}
\label{sec:setup}

We consider $D$-dimensional gravitational theory coupled to dilaton $\phi$ and $(n_A+1)$-form gauge field ($A=1, \ldots, m$). The action is given by
\begin{align}
S=\frac{1}{16\pi G_D} \int d^Dx \sqrt{-g} \left[ R - \frac{1}{2} \left( \partial \phi \right)^2
-\sum_{A=1}^m \frac{1}{2(n_A+2)!} e^{a_A \phi} F_{n_A+2}^2
\right].
\end{align}
Here $a_A$ is taken as
\begin{align}
a_A^2=4- \frac{2(n_A+1)(D-n_A-3)}{D-2}
\end{align}
in order to have asymptotically flat solutions, and so that extremal $q_A$-brane solutions obey a `no-force'
condition when the branes are static and parallel \footnote{ 
In this paper, we discuss the general $D$ cases.
When $D=10$, all branes in type IIA and IIB superstring theory can be discussed in this setup. Then the $D<10$ cases obviously include the compactified systems of these branes in superstring theory.
When $D=11$, M2- and M5-branes in M-theory can be discussed in this setup with no dilaton, $a_A=0$.
When $D=12$, we expect the branes in F-theory can be discussed.
Although F-theory seems a little strange in that it has two time-like directions
and the dilaton coupling $a_A$ is pure imaginary \cite{Khviengia:1997rh},
if one of time-like directions is Euclideanized or all branes wind around this time-like direction, the following discussions seem to be applicable.
However, at this stage, we cannot judge whether general brane systems in F-theory can be applied or not. It would be an interesting future work.}.

We compactify $(D-{d}-3)$-dimensional space as a rectangular torus $T^{D-{d}-3}$ and define its volume as $V_T$.
To make intersecting brane system in this setup, we put various branes on the torus.
We set $N_A$ $q_A$-branes 
winding around $q_A$ cycles of the torus ($A=1,\ldots, M$)  and assume that the branes are smeared along the other cycles\footnote{\label{ftnt-GL}
If there are cycles around which no brane wind, Gregory-Laflamme transitions occur on these cycles and the black brane is localized on these cycles, when the size of the event horizon of the black brane is smaller than the size of the cycles. 
In such cases, we cannot assume that the branes are smeared on these cycles and the effective action for the branes (\ref{gen-effective-action}) is modified as argued in \cite{Morita:2014ypa}.
By using this modified effective action, we can investigate the Gregory-Laflamme transitions in terms of the interacting separated branes. 
}. 
When we set the branes, we respect the intersection rule: $q_A$-brane and $q_B$-brane wind around the same $\bar{q}$ cycles of the torus, where $\bar{q}$ is determined as
\begin{align}
\bar{q}=\frac{(q_A+1)(q_B+1)}{D-2}-1-\frac{1}{2}\epsilon_A a_A \epsilon_B a_B
\label{intersection-rule} 
\end{align}
where $\epsilon_A$ is 1 ($-1$), if $q_A$-brane is electrically (magnetically) coupled to the field strength $F_{n_A+2}$ with $q_A=n_A$ ($q_A=D-n_A-4$). 
If all the branes satisfy this rule each other, `no-force' work among them when they are static.
Also we constrain ${d} \ge 1$ so that the co-dimension of the branes is higher than two.
(See Table \ref{table-branes} as an example.)

We will consider the black brane solution for this setup and argue that their thermodynamical properties can be 
explained via the microscopic theory of interacting branes (up to numerical factors).
It will work in general intersecting brane system as far as the branes satisfy the intersection rule.
However, 
discussion on the general systems is rather complicated, and hence we first demonstrate it for a two brane system.

\section{Example: two brane system}
\label{sec:twobrane}

\begin{table}
\begin{eqnarray*}
\begin{array}{l|c|c|c|c|c|c|c|c|}
& t & 1 & \cdots & {d}+2 & T^{q_1-\bar{q}} & T^{q_2-\bar{q}} & T^{\bar{q}} & T^{r}  \\
\hline
N_1~q_1\text{-brane} & - & && &- && - &   \\ \hline
N_2~q_2\text{-brane}& - & && & &-& - &  \\ \hline
\end{array}
\end{eqnarray*}
\caption{The brane configuration of intersecting two brane system.
Here $r= D- {d}-3 -(q_1+q_2-\bar{q})$ and $\bar{q}$ is fixed by (\ref{intersection-rule}).
$N_1$ $q_1$-branes wind around $q_1$ cycles of $T^{D-{d}-3}$ and  are smeared on the other directions of $T^{D-{d}-3}$. 
$N_2$ $q_2$-branes wind around $q_2$ cycles and $\bar{q}$ of them are the same cycles to the $q_1$-branes. 
}
\label{table-branes}
\end{table}

We consider the intersecting brane system which consists of $N_1$ $q_1$-branes and $N_2$ $q_2$-branes.
They wind around $q_1$ and $q_2$ cycles of the torus, 
sharing $\bar{q}$ cycles so that they satisfy the intersecting rule (\ref{intersection-rule}).
Note that, since $\bar{q}$ has to be a non-negative 
integer, $q_A$ and $D$ are restricted 
to particular combinations of values.

If we focus on only the non-compact $({d}+3)$-dimensional spacetime, the $q_A$-branes behave as BPS particles with mass $m_A \equiv \mu_A V_A$ ($A=1,2$).
Here $\mu_A$ is the tension of the single $q_A$-brane and $V_A$ is the volume of the $q_A$-dimensional torus which the brane winds around.
Due to the intersecting rule, no forces work between the branes when they are static.
However, if they are moving, the interactions arise.
Our proposal is that these interactions confine the branes 
and make them compose a bound state, and 
 this bound state explains the thermodynamics of the intersecting $q_1$-$q_2$ black hole.
To see this, we estimate the low energy effective action of this interacting brane system.
If the branes are well separated, the gravitational interactions dominate and the effective action for this system has the following structure
\begin{align}
S_{\text{eff}} = \int dt \left(
L_\text{kin}+L_\text{1-grav}+L_\text{2-grav}+L_\text{3-grav}+\cdots \right).
\label{moduli-D1D5}
\end{align}
The derivation of this effective action is summarized in \ref{app:eff}.
Here
\begin{align}
L_\text{kin}=
\sum_{i=1}^{N_1} \left( \frac{m_1}{2} \vect{v}{2}_i+ \frac{m_1}{8} (\vect{v}{2}_i)^2+ \cdots \right)
+\sum_{i=1}^{N_2} \left( \frac{m_2}{2} \vect{v}{2}_i+ \frac{m_2}{8} (\vect{v}{2}_i)^2 + \cdots \right)
\label{eff-kin}
\end{align}
is the kinetic term of the branes.
 $\vec v_i \equiv \partial_t \vec r_i$ and $\vec r_i$ are the velocity and the position of the $i$-th brane in the non-compact $({d}+2)$-dimensional space.
 We have assumed that the velocity is low ($|\vec v_i| \ll 1$) at low energy regime and used the non-relativistic approximation. We will justify this approximation soon.
In addition, we have assumed that the volume of the torus is small and the motions of the branes depend on time $t$ only\footnote{\label{ftnt-KK}
The small volume assumption is not essential in the following calculations, and
we apply it only to make the equations simpler.
}.

$L_\text{1-grav}$ is the interaction which arises from a single graviton (, gauge and dilaton) exchange between two branes,
\begin{align}
L_\text{1-grav}= 
\sum_{i\neq j}^{N_1}  \frac{\kappa^2_{{d}+3} m_1^2}{8{d}\Omega_{{d}+1}} \frac{\vect{v}{4}_{ij}}{\vect{r}{{d}}_{ij}}
+
\sum_{i\neq j}^{N_2}  \frac{ \kappa^2_{{d}+3} m_2^2}{8{d}\Omega_{{d}+1}} \frac{\vect{v}{4}_{ij}}{\vect{r}{{d}}_{ij}}
+\sum_{i=1}^{N_1} \sum_{j=1}^{N_2} \frac{ \kappa^2_{{d}+3} m_1 m_2}{{d}\Omega_{{d}+1}} \frac{\vect{v}{2}_{ij}}{\vect{r}{{d}}_{ij}}.
\label{eff-2-body-D1-D5}
\end{align}
Here $\vec{r}_{ij}\equiv \vec{r}_i-\vec{r}_j$ and $\vec{v}_{ij}\equiv \vec{v}_i-\vec{v}_j$ denote the relative position and the relative velocity of the $i$-th and $j$-th branes, and $\kappa^2_{{d}+3} \equiv  \kappa_D^2/V_T $ is the $({d}+3)$-dimensional gravitational coupling constant.
$\Omega_{ d+1}\equiv 2\pi^{\frac{ d}{2}+1}/\Gamma(\frac{ d}{2}+1)$ is the volume of a unit $( d+1)$-sphere.
The first and second terms describe the two-body interactions between two $q_1$-branes and two $q_2$-branes, respectively.
The third term is the two-body interactions between a $q_1$-brane and a $q_2$-brane.
There are higher order terms of $\vec{v}_{ij}$ but we have omitted them in this equation, since  $|\vec{v}_{ij}|$ would be small in the low energy regime.
Note that the power of $\vec{v}_{ij}$ of the third term is lower than the others, and
it implies that the third term would dominate.
Hence we treat this term separately and define it as 
\begin{align}
L_1 \equiv \sum_{i=1}^{N_1} \sum_{j=1}^{N_2} \frac{ \kappa^2_{{d}+3} m_1 m_2}{{d}\Omega_{{d}+1}} \frac{\vect{v}{2}_{ij}}{\vect{r}{{d}}_{ij}}.
\end{align}
We will soon see that it indeed becomes relevant.

Similarly the effective action has various interaction terms through the multi-graviton exchanges.
Here we write down only the terms proportional to the lowest power of $v$, since they will become relevant at low energy.
Among 3-graviton exchange interactions, the following term will be relevant, 
\begin{align}
L_\text{3-grav} \ni  L_2 \sim
\sum_{i=1}^{N_1} \sum_{j=1}^{N_1} \sum_{k=1}^{N_2} \sum_{l=1}^{N_2}
\frac{\kappa^6_{{d}+3}  m_1^2 m_2^2}{\Omega_{{d}+1}^3} \left(  \frac{\vect{v}{4}_{ij}}{\vect{r}{{d}}_{ij} \vect{r}{{d}}_{ik} \vect{r}{{d}}_{il}  }+ \cdots \right).
\label{eff-4-body}
\end{align}
We define this term as $L_2$.
`$\sim$' in this article denotes equality not only including dependence on physical parameters but also including all factors of $\pi$.
When we derive these interactions in \ref{app:eff}, we do not fix the precise numerical coefficients of these interactions.
Since  we will consider an order estimate for the thermodynamics of this interacting brane system, the precise expressions are not important. 
Similarly the system has the following interactions
\begin{align}
L_{n} \sim & \sum_{i_1, \dots, i_n}^{N_1}\sum_{j_1, \dots, j_n}^{N_2} 
\left( \kappa_{{d}+3}^{2(2n-1)} \frac{m_1^n m_2^n }{\Omega_{{d}+1}^{2n-1}} \prod_{k=2}^n \prod_{l=1}^n
\frac{ 1 }{  \vect{r}{{d}}_{i_1 i_k} \vect{r}{{d}}_{i_1 j_l} } \vect{v}{2n} + \cdots \right),
\label{moduli-multi-graviton} 
\end{align}
which describes the $2n-1$ graviton exchange among $n$ $q_1$-branes and $n$ $q_2$-branes.

From now, we estimate the dynamics of this system by using the virial theorem.
We first assume that the branes are confined due to the interactions, and the branes satisfy
\begin{align}
\vec{v}_{ij} \sim  v, \qquad \vec{r}_{ij} \sim r.
\label{assumption-scales}
\end{align}
Here $v$ and $r$ are the characteristic scales of the velocity and size of the branes in 
the bound state which do not depend on the species of the branes.
(Note that since the masses of the $q_1$-brane and $q_2$-brane are generally different, we naively expect that  these scales should depend on the species of the branes.
However we will soon see that it does not occur in the bound state.)
Then we can estimate the scales of the terms (\ref{eff-kin}), (\ref{eff-2-body-D1-D5}) and (\ref{eff-4-body}) in the effective Lagrangian as
\begin{align}
L_\text{kin} \sim & N_1 m_1 v^2+ N_1 m_1 v^4+ \cdots + N_2 m_2 v^2+ N_2 m_2 v^4  + \cdots, \nonumber \\
L_\text{1-grav} \sim& \frac{ \kappa^2_{{d}+3} N_1^2 m_1^2}{\Omega_{{d}+1}} \frac{v^4}{r^{{d}}}+\frac{ \kappa^2_{{d}+3} N_2^2 m_2^2}{\Omega_{{d}+1}} \frac{v^4}{r^{{d}}} + \frac{ \kappa^2_{{d}+3} N_1 N_2 m_1 m_2}{\Omega_{{d}+1}} \frac{v^2}{r^{{d}}} + \cdots, \nonumber \\
L_\text{3-grav} \sim &
\frac{ \kappa^6_{{d}+3} N_1^2 N_2^2 m_1^2 m_2^2}{\Omega_{{d}+1}^3} \frac{v^4}{r^{3{d}}} + \cdots.
\label{estimate-D1D5}
\end{align}
The Lagrangian also have other terms (\ref{moduli-multi-graviton}) but we will consider them later.
Here we consider which terms in (\ref{estimate-D1D5}) dominate at the low energy where $v$ would be small ($v \ll 1 $). 
In the second line of (\ref{estimate-D1D5}), the third term which is from $L_1$ (\ref{eff-2-body-D1-D5}) would dominate, since the power of $v$ is lowest \footnote{\label{ftnt-reduction} If the numbers and masses of the branes are quite different, e.g. $m_1 N_1 v^2  \ll m_2 N_2 $, we can ignore the contribution from the another species of the branes and
 the system would reduce to the single-species brane system with no intersecting which has been studied in \cite{Morita:2013wfa}. 
 }.
Suppose that this term is balanced to the term in the third line  which is from $L_2$ (\ref{eff-4-body}) due to the virial theorem, 
we obtain the relation between $v$ and $r$ as
\begin{align}
\frac{ \kappa^2_{{d}+3} N_1 N_2 m_1 m_2}{\Omega_{{d}+1}} \frac{v^2}{r^{{d}}}
 \sim 
\frac{ \kappa^6_{{d}+3} N_1^2 N_2^2 m_1^2 m_2^2}{\Omega_{{d}+1}^3} \frac{v^4}{r^{3{d}}} \quad \Longrightarrow \quad  
v^2 \sim  \frac{\Omega_{{d}+1}^2r^{2{d}}}{ \kappa^4_{{d}+3} N_1 N_2 m_1 m_2}.
\label{scale-v-r}
\end{align}
Now we see that the other terms listed in (\ref{estimate-D1D5}) are indeed subdominant at this scaling.
To see this, it is convenient to define $\cQ_A \equiv \kappa^2_{{d}+3} m_A N_A/\Omega_{{d}+1}$.
Then the scaling relation (\ref{scale-v-r}) is rewritten as $v^2 \sim r^{2 {d}}/\cQ_1 \cQ_2$ and the low energy $|v| \ll 1$ indicates $r^d/\cQ_1$ and $r^d/\cQ_2$ are small\footnote{More precisely speaking, $|v| \ll 1$ indicates that $r^{2d}/\cQ_1\cQ_2 \ll 1$ and does not indicate that $r^d/\cQ_1$ and $r^d/\cQ_2$ are both small.
For example, if $\cQ_1/r^d \ll \cQ_2/r^d$, $r^d/\cQ_2$ can be large. However this is the situation that we should ignore another species of the brane as we argued in footnote \ref{ftnt-reduction}, and we can exclude this possibility. Hence we can regard that  $r^d/\cQ_1$ and $r^d/\cQ_2$ are both small. }.
Thus the terms in the Lagrangian (\ref{estimate-D1D5}) scale as,
\begin{align}
L_\text{kin}  \sim & \frac{ \Omega_{{d}+1} r^{{d}}}{\kappa^2_{{d}+3}} \frac{r^d}{\cQ_2} + \frac{ \Omega_{{d}+1} r^{{d}}}{\kappa^2_{{d}+3}} \frac{r^d}{\cQ_2}  \frac{r^{2d}}{\cQ_1 \cQ_2} + \cdots +  \frac{ \Omega_{{d}+1} r^{{d}}}{\kappa^2_{{d}+3}}  \frac{r^{d}}{\cQ_1 }  +   \frac{ \Omega_{{d}+1} r^{{d}}}{\kappa^2_{{d}+3}} \frac{r^{d}}{\cQ_1 }  \frac{r^{2d}}{\cQ_1 \cQ_2} + \cdots,  \nonumber \\
L_\text{1-grav} \sim&   \frac{ \Omega_{{d}+1} r^{{d}}}{\kappa^2_{{d}+3}} \left( \frac{r^{d}}{\cQ_2 }\right)^2
+  \frac{ \Omega_{{d}+1} r^{{d}}}{\kappa^2_{{d}+3}} \left( \frac{r^{d}}{\cQ_1 }\right)^2+  \frac{ \Omega_{{d}+1} r^{{d}}}{\kappa^2_{{d}+3}} + \cdots, \nonumber \\
L_\text{3-grav} \sim &   \frac{ \Omega_{{d}+1} r^{{d}}}{\kappa^2_{{d}+3}} + \cdots.
\label{r-Q}
\end{align}
Here the ordering of the terms is the same as in eq.\,(\ref{estimate-D1D5}).
We see that $L_1$ and $L_2$ scale as $  \Omega_{{d}+1} r^{{d}}/\kappa^2_{{d}+3} $, while the other terms earn the factors of $r^{d}/\cQ_1 $ and/or $r^{d}/\cQ_2 $ and are suppressed at low energy ($r^d \ll \cQ_1, \cQ_2$).
This self-consistently ensures our assumption that $L_1$ and $L_2$ are relevant at the low energy, and hence the scaling relation (\ref{scale-v-r}) is confirmed.
Note that the masses of the branes always appear as the combination $m_1 m_2$ in $L_1$ and $L_2$, and it ensures that the scales of the position $r$ and velocity $v$ are independent of the species of the branes as we assumed in (\ref{assumption-scales}).

So far we have considered the terms up to $L_2$, and now we consider $L_n$ (\ref{moduli-multi-graviton}) ($n \ge 3$) too. 
We can see that at the scaling (\ref{scale-v-r}) which was derived via the virial theorem $L_1 \sim L_2 $, all the other interactions $L_n$ also become the same order.
It means that the branes are strongly coupled in the bound state.
We called such a bound state as `warm $p$-soup' in Ref.~\cite{Morita:2013wfa}.

From now, we evaluate the thermodynamical quantities of the bound state.
By substituting the relation (\ref{scale-v-r}) to the Lagrangian $L \sim L_1$, we estimate the free energy of the system as
\begin{align}
F \sim L_1 \sim \frac{ \Omega_{{d}+1} r^{{d}}}{\kappa^2_{{d}+3}}.
\label{free-energy-D1D5} 
\end{align}
Here we consider temperature dependence.
When the bound state is thermalized, we treat $\vec{r}_i$ as a thermal field (particle) and expand $\vec{r}_i(t) = \sum_n \vec{r}_{i(n)} \exp\left(i \frac{2 \pi n}{\beta}t  \right) $. 
Hence we assume that the velocity $v = \partial_t r$ are characterized by the temperature of the system through
\begin{align}
v  \sim \pi T r.
\label{key-assumption}
\end{align}
Note that such a relation is not held generally if the system has a mass gap, but
there would be no mass gap in the interacting brane system as argued in Ref.~\cite{Morita:2013wfa}.
Then, from (\ref{scale-v-r}), we obtain the relation between the size of the bound state and the temperature
\begin{align}
r \sim \left(   \frac{ \pi^2 \kappa^4_{{d}+3} N_1 N_2 m_1 m_2 T^2 }{\Omega_{{d}+1}^2} \right)^{\frac{1}{2 ({d}-1)}} .
 \label{scale-r-T}
\end{align}
Here we have assumed ${d} \neq 1$, and we will consider ${d}=1$ case later.
By substituting this relation into the free energy (\ref{free-energy-D1D5}), 
we estimate the entropy of the bound state as
\begin{align}
&S_{\text{entropy}} = - \frac{\partial F}{\partial T}  \sim
\left(   \pi^2 N_1 N_2 m_1 m_2 \right)^{\frac{{d}}{2 ({d}-1)}}
\left(   \frac{ \kappa^2_{{d}+3}   T }{\Omega_{{d}+1}} \right)^{\frac{1}{ {d}-1}}.
\label{entropy-D1D5}  
\end{align}

We compare the obtained quantities with the corresponding black hole solution.
In the near extremal regime, the black hole thermodynamics tells us, 
\begin{align}
F&= -
\frac{d-1}{2}\frac{\Omega_{{d}+1} r_H^{{d}}}{\kappa^2_{{d}+3}},
\label{free-energy-D1D5-GR} \\
S_{\text{entropy}}&=
\left( \pi^2   N_1 N_2 m_1 m_2 \right)^{\frac{{d}}{2 ({d}-1)}}
\left(   \frac{2^{2 d+1} \kappa^2_{{d}+3}   T }{{ d}^{ d+1} \Omega_{{d}+1}} \right)^{\frac{1}{ {d}-1}},
\label{entropy-D1D5-GR} \\
r_H&= \left(   \frac{ 64 \pi^2 \kappa^4_{{d}+3} N_1 N_2 m_1 m_2 T^2 }{{ d}^4\Omega_{{d}+1}^2} \right)^{\frac{1}{2 ({d}-1)}} .
\label{eq-tmp-horizon}
\end{align}
The derivation of these expressions is summarized in \ref{app:thermo}, and we have taken $M=2$.
Here $r_H$ is the location of the horizon.
Therefore, if we identify the size of the bound state $r$ with the horizon $r_H$, our result (\ref{free-energy-D1D5}), (\ref{scale-r-T}) and (\ref{entropy-D1D5}) reproduce the parameter dependences of the black hole thermodynamics including $\pi$.
($r_H$ depends on the coordinate and we have argued what coordinate is natural in \cite{Morita:2013wfa}.)
This agreement may indicate that 
the interacting $q_1$ and $q_2$-branes described by the effective action (\ref{moduli-D1D5}) provide
the microscopic origin of the $q_1$-$q_2$ black hole thermodynamics.

Now we comment on the assumption $r^d \ll \cQ_1, \cQ_2$ which we have used when we consider the effective action (\ref{moduli-D1D5}).
At the scale (\ref{scale-r-T}), this relation becomes
\ba
T \ll \frac{(\cQ_1)^{\frac12-\frac1d}}{\pi (\cQ_2)^{\frac12}},\quad
\frac{(\cQ_2)^{\frac12-\frac1d}}{\pi (\cQ_1)^{\frac12}}.
\ea
Since we consider here the situation $\cQ_1/\cQ_2\sim \cO(1)$, this means 
$T \ll 1/(\cQ_1)^\frac1d, 1/(\cQ_2)^\frac1d$ and this is the near extremal limit in supergravity. 
Thus our analysis is valid when we consider the near extremal black holes.

Finally we consider the ${d}=1$ case.
In this case the relations (\ref{scale-v-r}) and (\ref{key-assumption}) fix  $T$ as
\begin{align}
T \sim \left( \frac{\Omega_{2}^2}{ \kappa^4_{4} N_1 N_2 m_1 m_2 \pi^2 } \right)^{\frac{1}{2 }} .
\end{align}
and $r$ remains as a free parameter.
This is the Hagedorn behavior, and this temperature is coincident with the Hagedorn temperature in supergravity (\ref{HagedornT}).
Thus the $p$-soup proposal works even in the $d=1$ case too.

\section{General intersecting black brane}
\label{sec:general}

We discuss general black brane system with arbitrary number of species of branes. 
We can also introduce momentum along one of the cycles of the torus if all the branes wind around this cycle. (We define $R$ as the radius of this cycle.)
In this case, the momentum is quantized as $N_P/R$ and there are $N_P$ gravitational waves each of which carries a momentum $1/R$.
We can regard this gravitational wave as a 0-brane with mass $1/R$ via the KK reduction of this cycle.
Although the setup is complicated, we will see a simple result that the effective theory of the separated intersecting branes explains the black hole thermodynamics of the corresponding black brane at low energy.

In \ref{app:eff}, we argue the derivation of the low energy effective action for $N_A$ $q_A$-branes ($A=1,\ldots, M$) where $M$ is the number of the species of the charges (including momentum if we introduce it).
The dominant terms of the action are given by
\begin{align}
\label{lag-gen}
S_{\text{eff}} =& \int dt \left(L_1+L_2+\cdots\right), \\
L_n \sim &
\sum_{i_1,\ldots,i_n}^{N_1}\sum_{j_1,\ldots,j_n}^{N_2}\cdots \sum_{k_1,\ldots,k_n}^{N_M}
\frac{\kappa_{d+3}^{2(nM-1)}\,\hat\prod_A\,m_A^n}{\Omega_{d+1}^{nM-1}}
\left(\vect v{2n}\prod_{a=2}^n\prod_{b=1}^n\cdots\prod_{c=1}^n\frac{1}{\vect rd_{i_1i_a}\vect rd_{i_1j_b}\cdots \vect rd_{i_1k_c}}+\cdots\right),
\end{align}
where $\hat\prod_A$ is the product including momentum.
Similar to the two brane case, we apply the virial theorem and estimate the thermodynamics of this system.
With the assumption (\ref{assumption-scales}), the first two terms of the action scale as 
\ba
L_1 \sim \frac{\kappa_{d+3}^{2(M-1)}}{\Omega_{d+1}^{M-1}} \left(\hat{\prod_A}\, m_A N_A \right) \frac{v^2}{r^{d(M-1)}} 
\,,\quad
L_2 \sim \frac{\kappa_{d+3}^{2(2M-1)}}{\Omega_{d+1}^{2M-1}} \left(\hat{\prod_A}\, m_A N_A \right)^2 \frac{v^4}{r^{d(2M-1)}} 
\,,
\label{L1L2-gen}
\ea
Therefore through the virial theorem $L_1\sim L_2$, we obtain the relation between $v$ and $r$ as
\ba
v^2  \sim \frac{\Omega_{d+1}^M r^{dM}}{\kappa_{d+3}^{2M}\,\hat\prod_A\, N_A m_A}.
\ea
Then the free energy can be estimated as
\ba
F\sim L_1 \sim \frac{\Omega_{d+1}r^d}{\kappa_{d+3}^2}\,.
\ea
Note that this expression is common for all $M$.
Moreover, if we assume the relation between the velocity and temperature of the system $v\sim \pi Tr$, we can estimate the radius of horizon
\ba
r \sim \left(\frac{\pi^2 T^2 \kappa_{d+3}^{2M} \,\hat\prod_A\, N_A m_A}{\Omega_{d+1}^M}\right)^{\frac{1}{dM-2}}\,.
\ea
for $dM-2 \neq 0$.
Using the expressions of free energy and horizon radius, the entropy can be written as 
\ba
S_{\text{entropy}}=-\frac{\partial F}{\partial T} 
\sim \left(\pi^2\,\hat{\prod_A}\, N_Am_A\right)^{\frac{d}{dM-2}}\left(\frac{\kappa^2_{d+3}}{\Omega_{d+1}}\right)^{\frac{2}{dM-2}} T^{\frac{2d}{dM-2}-1}\,.
\ea

Now we compare our results with the corresponding black hole thermodynamics.
The results in the black hole are shown in eqs.\,(\ref{rH:sugra}), (\ref{en:sugra}) and (\ref{F:sugra}).
We can check that our results are coincident with the black hole thermodynamics up to rational numerical factors.
Similar consistency can be also seen for the Hagedorn case ($dM-2=0$).

\section{Discussions}
\label{sec-conclusion}

We generalized the $p$-soup proposal \cite{Morita:2013wfa} to the intersecting brane systems.
We figured out that, in case the branes satisfy the intersection rule (\ref{intersection-rule}), the bound state of the branes exhibits the thermodynamical properties which agree with the corresponding black brane.
Although the intersecting brane systems are complicated themselves, this result is strikingly simple, and we believe that our proposal captures profound natures of the black hole microstates.

Also we can apply the results to superstring theory and M-theory.
For example, the D1-D5 brane system studied in \cite{Morita:2014cfa} can be mapped to other brane configurations such as D0-D4 system or M5-P system via string dualities \cite{Martinec:1999sa}.
Then we can investigate the microstates of these branes similarly.
Furthermore the phase transitions between these branes \cite{Martinec:1999sa} could be understood microscopically in a fashion of \cite{Morita:2014ypa}.
In such a way, we can study the various black brane dynamics in string theory through our proposal.

Moreover, our discussion would be important to investigate supersymmetric gauge theories. 
In the case of the single-species D$p$/M-branes, through the gauge/gravity correspondence, the duality between the black branes and the supersymmetric gauge theories on the branes at finite temperature is expected \cite{Itzhaki:1998dd}.
We can explain this duality by using the fact that the effective action for the interacting $N$ branes (a variety of (\ref{single-brane})), which is obtained from gravity, is also derived from the supersymmetric gauge theory on the branes as a low energy effective action \cite{Morita:2013wfa}.
Not only that, the effective action might be useful to estimate several quantities in the gauge theory which cannot be calculated in the gravity \cite{Morita:2014ypa}.
In the case of the intersecting brane systems, much more varieties of supersymmetric gauge theories 
appear on the branes than the single-species brane case.
There, the effective action (\ref{lag-gen}) may play a key role to understand the dynamics of these gauge theories.

In this way, the $p$-soup proposal may be important in the various contexts in theoretical physics. In order to establish the proposal, we have to derive the thermodynamical quantities from the effective action (\ref{lag-gen}) exactly and compare them with the black hole results. 
However even fixing the coefficients of the interactions in the effective action (\ref{lag-gen}) is difficult in the intersecting brane system, and we need to find a clever way to solve this issue.
If we could achieve it, it must be a valuable step toward the understanding of the black hole microstate.

\subsection*{Acknowledgments}
We would like to thank Andrew Hickling, Toby Wiseman and Benjamin Withers for useful discussions through the collaboration.
We also thank Nobuyoshi Ohta for useful comments.
The work of T.~M. is supported in part by Grant-in-Aid for Scientific Research (No.\,15K17643) from JSPS.

\appendix

\section{Thermodynamics of intersecting black brane}
\label{app:thermo}

In this appendix, we introduce the intersecting black brane system and its thermodynamical properties.
We consider the brane configuration discussed around equation (\ref{intersection-rule}).
In the non-extremal case, the metric for the intersecting black brane with momentum is given by
\begin{align}
ds_D^2 =& \prod_A H_A^{\frac{q_A+1}{D-2}}\left[
-\prod_A H_A^{-1} K^{-1} f dt^2
+\prod_A H_A^{-1} K d\hat y_1^2
\right.\nt &\qquad\qquad\qquad \left.
+\sum_{\alpha=2}^{D-d-3} \prod_A H_A^{-\delta_A^{(\alpha)}} dy_\alpha^2
+f^{-1}dr^2+r^2d\Omega_{d+1}^2
\right],
\label{metric-gen}
\\
&H_A = 1+\frac{Q_A}{r^d}\,,\quad
f= 1-\frac{2\mu}{r^d}\,.
\label{H}
\end{align}
where the index $A$ is given for each species of branes and $y_\alpha$ ($\alpha=1,\ldots, D-d-3$) denote the coordinates for $T^{D-d-3}$.
Here we have introduced momentum for one of the isometric directions $y_1$ around which all the branes wind. Then the $y_1$ direction is Lorentz boosted:
\ba
K = 1+\frac{P}{r^d}\,,\quad
d\hat y_1 = dy_1 - \frac{\hat K}{K} dt\,;\quad
\hat K=\frac{\sqrt{P(P+2\mu)}}{r^d}\,.
\label{K}
\ea
For the other directions of $y_\alpha$, some of the branes wind around them.
If the brane $A$ wind around the $y_\alpha$-cycle, $\delta_A^{(\alpha)}=1$.
If not, $\delta_A^{(\alpha)}=0$.
(If we don't introduce momentum $P=0$, not all the branes have to wind around the $y^1$-cycle and we treat $y^1$ as one of $y^\alpha$'s.)

The gauge potentials and dilaton are given by
\begin{align}
E_A =\sqrt{\frac{Q_A}{(Q_A+2\mu)}}\frac{f}{H_A}\,,\quad e^{\phi}= \prod_A H_A^{\frac{\epsilon_A a_A}{2}}.
\label{gauge}
\end{align}
Thus the field strength behaves $F_A = \partial_r E_A \to d \sqrt{Q_A(Q_A+2\mu)}/r^{d+1}$ ($r\to \infty$), and we can read off the charges of the black brane from it. 
Then by dividing the charges by $\sqrt{2}\kappa_{D}\mu_A$ where $\mu_A$ is the tension of a single $q_A$-brane, we obtain the number of the $q_A$-branes which compose the black brane as
\begin{align}
N_A= \frac{{d}\Omega_{{d} +1} }{2 \kappa_D^2 \mu_A } \sqrt{Q_A(Q_A+2\mu)} \frac{V_T}{V_A}.
\label{brane-number}
\end{align}
Here $V_T$ is the volume of $T^{D-{d}-3}$ and $V_A$ is the volume of the $T^{q_A}$ around which the $q_A$-branes wind.
(The volume dependence has arisen since the branes are smeared on the $(D-d-3-q_A)$-dimensional torus with the volume $V_T/V_A$.)
Hence $Q_A$ is written as 
\begin{align}
Q_A = \cQ_A - \mu + \cdots, \qquad \cQ_A=  \frac{2 \kappa_D^2 \mu_A V_A}{{d}\Omega_{{d} +1} V_T} N_A
\end{align}
in the near extremal regime ($Q_A \gg 2\mu$).
It is convenient to define the $({d}+3)$-dimensional gravitational coupling $\kappa_{{d}+3}^2 \equiv \kappa^2_{D}/V_T $ and mass of the $q_A$-brane $m_A \equiv \mu_A V_A$.
By using them, $\cQ_A$ is rewritten as
\begin{align}
\cQ_A = \frac{2 \kappa_{d+3}^2 m_A}{{d}\Omega_{{d} +1} } N_A.
\label{Q-N-relation}
\end{align}

The momentum is also quantized because of compactification.
Since $\hat{K}/K$ can be regarded as a Kaluza-Klein gauge potential,  similar to (\ref{brane-number}), we can easily read off the relation between the parameter $P$ and $N_P$ which is the number of the momentum with the unit $1/R$ where $R$ is the radius of the $y^1$-cycle,
\ba
P=\cP- \mu + \cdots,\qquad
  \cP=\frac{2\kappa_{d+3}^2}{d\Omega_{d+1}R}N_P.
\ea

We discuss the thermodynamical quantities of this black brane in the near extremal regime.
To do it, we formally reduce the torus part and regard the black brane as a $(d+3)$-dimensional black hole.
Through the dimensional reduction, the $(d+3)$-dimensional metric in Einstein frame becomes
\ba
ds_{d+3}^2 = -\prod_A H_A^{-\frac{d}{d+1}} K^{-\frac{d}{d+1}} f dt^2 
+ \prod_A H_A^{\frac{1}{d+1}} K^{\frac{1}{d+1}} \left(f^{-1}dr^2+r^2d\Omega_{d+1}^2\right).
\ea
Then we can read off the ADM energy $E$ from the $g_{tt}$ component of the metric,
\begin{align}
E=\frac{\Omega_{d+1}}{\kappa^2_{d+3}}\left( d+1 - \frac{dM}{2} \right)\mu + \hat{\sum_A} m_A N_A
\label{ADM}
\end{align}
where $\hat\sum_A$ is the sum including momentum as one of charges as $m_A N_A=N_P/R$ if we introduce it (i.e. $P\neq 0$) and $M$ is the number of the species of the charges (including momentum).

The Hawking temperature is obtained through the smoothness condition of the metric in the Euclidean signature.
In the near extremal limit $\mu\to 0$, we obtain
\ba
T_H 
= \lim_{\mu\to 0} \frac{1}{4\pi}\frac{d}{(2\mu)^{1/d}}\,\hat{\prod_A} \left(\frac{2\mu}{2\mu+Q_A}\right)^{\frac12}
= \frac{d}{4\pi}(2\mu)^{\frac{M}{2}-\frac{1}{d}}\left(\hat{\prod_A}\, {\cQ_A}\right)^{-1/2}
\label{TH:sugra}
\ea
where $\hat\prod_A$ is the product including momentum if we introduce it.
Then, if $M\neq 2/d$, we see the radius of horizon can be expressed as
\ba
r_H = (2\mu)^{1/d} 
= \left(\frac{16\pi^2}{d^2}T_H^2 \,\hat{\prod_A}\, \cQ_A\right)^\frac{1}{dM-2}
= \left(\frac{2^{M+4}\pi^2T_H^2 \kappa^{2M} \,\hat\prod_A\, N_A m_A}{d^{M+2}\Omega_{d+1}^M}\right)^\frac{1}{dM-2}.
\label{rH:sugra}
\ea
If $M= 2/d$, the radius of the horizon remains as a free parameter of the system while temperature is fixed as
\begin{align}
T_H= \frac{d}{4\pi}\left(\hat{\prod_A}\, {\cQ_A}\right)^{-1/2}.
\label{HagedornT}
\end{align}
This is called the Hagedorn behavior and the thermodynamical quantities are characterized by $r_H$ rather than temperature.

The area of horizon can be calculated using the angular part ($d\Omega_{d+1}^2$) of the metric, and it is given by
\ba
A_{ d+1} 
= \Omega_{d+1}\left[r^2\,\hat{\prod_A}\, H_A^\frac{1}{d+1} K^\frac{1}{d+1}\right]^{\frac{d+1}{2}}_{r=r_H}
= 
\Omega_{d+1} (2\mu)^{\frac{d+1}{d}-\frac{M}{2}}\,\hat{\prod_A}\, (Q_A+2\mu)^{1/2}.
\ea
Then we obtain the entropy 
by using Bekenstein-Hawking entropy formula
\ba
S_{ d+1} &=& \frac{2\pi}{\kappa_{d+3}^2}A_{ d+1}
= \frac{2\Omega_{d+1}}{\kappa_{d+3}^2}
\left(\frac{4T_H}{d}\right)^{\frac{2d}{dM-2}-1}
\left(\pi^2\,\hat{\prod_A}\, \cQ_A\right)^{\frac{d}{dM-2}}
\nt
&=& \left(\frac{4^{2d+1}\kappa_{d+3}^4}{d^{2(d+1)}\Omega_{d+1}^2}\right)^{\frac{1}{dM-2}}
\left(\pi^{2}\,\hat{\prod_A}\, N_Am_A\right)^{\frac{d}{dM-2}}
T_H^{\frac{2d}{dM-2}-1}.
\label{en:sugra}
\ea
Finally we obtain the free energy $F=E-TS$ from (\ref{ADM}), (\ref{TH:sugra}) and (\ref{en:sugra}), 
\ba
F = -\frac{dM-2}{4}
\frac{\Omega_{d+1}}{\kappa_{d+3}^2}
\left(\frac{16\pi^2}{d^2} T_H^2\,\hat{\prod_A}\, \cQ_A\right)^{\frac{d}{dM-2}}
= -\frac{dM-2}{4}\frac{\Omega_{d+1} r_H^d}{\kappa_{d+3}^2}
\label{F:sugra}
\ea
where we have omitted the temperature independent terms in (\ref{ADM}).

\section{Derivation of the effective action of intersecting branes}
\label{app:eff}

In this appendix, we derive the effective action for separated intersecting $N_A$ $q_A$-branes.
We assume that the branes satisfy the intersection rule (\ref{intersection-rule}), then
no force work between the branes if they are static.
When the branes are excited and moving, they gravitationally interact with each other.
We derive the effective action for this system at low energy.

However our purpose of this paper is the derivation of the parametric dependence of the thermodynamical quantities of this system and we do not need the precise numerical coefficients of the effective action.
Hence we employ the probe action and read off the interactions between the branes up to numerical factors (but we keep the $\pi$ dependences).

First we consider the extremal brane geometry
\begin{align}
ds_D^2 &= \prod_A H_A^{\frac{q_A+1}{D-2}}\left[
-\prod_A H_A^{-1} K^{-1}  dt^2
+\prod_A H_A^{-1} K d\hat y_1^2
+\sum_{\alpha=2}^{D-d-3} \prod_A H_A^{-\delta_A^{(\alpha)}} dy_\alpha^2
+\sum_{i=1}^{d+2}dx_i^2\right], \\
&H_A = 1+\frac{\cQ_A}{r^d}\,,\quad
K = 1+\frac{\cP}{r^d},
\label{H2}
\end{align}
where $y_\alpha$ ($\alpha=1,\ldots, D-d-3$) denote the coordinates for $T^{D-d-3}$ and  $x_i$ ($i=1,\ldots, d+2$) denote the coordinates for the non-compact $(d+2)$-dimensional space.
This geometry is merely obtained from the black brane (\ref{metric-gen}) with $\mu=0$.
To see the interactions, we choose one of the branes in the system and consider the probe brane action for this brane \cite{Morita:2013wfa}
\ba
S_{\text{probe},A}= -\mu_A \left(\int d^{q_A+1}\xi \sqrt{-\det\gamma_{\mu\nu}}+\int \hat E_A \right)
\ea
where $\gamma_{\mu\nu}$ is the worldvolume metric induced from the spacetime metric (\ref{metric-gen})
\ba
\gamma_{\mu\nu} = \partial_\mu Z^m \partial_\nu Z^n g_{mn} e^{-\frac{\epsilon_A a_A}{q_A+1}\phi}
\ea
and $\hat E_A$ is the pullback of gauge potential (\ref{gauge}) to the brane worldvolume.
We take the static gauge and assume that $Z^m$ depend only on time $t$, i.e. $Z^m=Z^m(t)$. (See footnote \ref{ftnt-KK}.) 
Also we ignore the motion of the branes on the torus by assuming the branes are smeared.
(See footnote \ref{ftnt-GL}.) 
Then the probe action becomes
\ba
S_{\text{probe},A=1}
= -m_1\int dt \left[
\frac{1}{H_1}\sqrt{1-\vect{v}{2}_1 K \prod_A H_A}
-\left(\frac{1}{H_1}-1\right)
\right]
\ea
where $m_1=\mu_A V_A$ is mass of the probe brane and $ v^m_1 = \partial_t Z^m$ ($m=1,\ldots, d+2$) is velocity in the non-compact $(d+2)$-dimensional space.

We assume the velocity is low ($v_1\ll 1$) and expand the action as
\ba
S_{\text{probe},A=1}
= m_1\int dt \left[
\frac{\vect{v}{2}_1}{2} K\prod_{A\neq 1} H_A
+\frac{(\vect{v}{2}_1)^2}{8} H_1 K^2\prod_{A\neq 1} H_A^2
+\cdots
\right].
\label{probe-gen-1}
\ea
Here we omit the rest mass term. By using eq.\,(\ref{H2}), we can rewrite this as
\begin{align}
S_{\text{probe},A=1}
=& m_1\int dt \left[
\frac{\vect{v}{2}_1}{2}\left(1+\frac{\sum_{A\neq 1}\cQ_A +\cP}{r^d}+\cdots
+\frac{\hat\prod_{A\neq 1}\,\cQ_A}{r^{d(M-1)}}\right)
\right.\nt&\left.
+\frac{(\vect{v}{2}_1)^2}{8}\left(1+\frac{\cQ_1+2(\sum_{A\neq 1}\cQ_A+\cP)}{r^d}+\cdots
+\frac{\cQ_1\,\hat\prod_{A\neq 1}\cQ_A^2}{r^{d(2M-1)}}\right)
+\cdots
\right]
\label{probe-gen}
\end{align}
where $M$ is the number of the species of charges including momentum.

Now we argue how to read off the interactions between the separated branes from this probe action.
We first consider $M=2$ case with no momentum ($P=0$) which we discuss in section \ref{sec:twobrane}.
In this case the probe action (\ref{probe-gen}) becomes
\begin{align}
S_{\text{probe},A=1}
=& m_1\int dt \left[
\frac{\vect{v}{2}_1}{2}\left(1+\frac{\cQ_2}{r^d}\right)
+\frac{(\vect{v}{2}_1)^2}{8}\left(1+\frac{\cQ_1+2\cQ_2}{r^d}
+\frac{2 \cQ_1 \cQ_2}{r^{2d}}+\frac{\cQ_1\cQ_2^2}{r^{3d}}\right)
+\cdots
\right].
\end{align}
In this action, since $\cQ_A$ involves $\kappa^2_{d+3}$ as shown in (\ref{Q-N-relation}), the terms proportional to $(\cQ_A)^n$ would describe the $n$-graviton (, dilaton and gauge field) exchange interaction between the probe brane and other branes.  
For example, the first term and the third term do not depend on $Q_A$ and they are merely the non-relativistic kinetic term $m_1 \vect{v}{2}_1$ and its relativistic correction $m_1(\vect{v}{2}_1)^2$, respectively.
On the other hand, the second term is proportional to $\cQ_2 \propto \kappa_{d+3}^2 m_2 N_2 $ and hence would describe the two body interaction between $q_1$ and $q_2$-brane.
Thus it indicates that the separated intersecting $N_1$ $q_1$-brane and $N_2$ $q_2$-brane system has interactions
\begin{align*}
\sum_{i=1}^{N_1}\sum_{j=1}^{N_2} \frac{\kappa_{d+3}^2 m_1m_2}{d\Omega_{d+1}}\frac{\vect  v2_{ij}}{\vect rd_{ij}}.
\end{align*}
Here $\vec r_{ij}=\vec r_i-\vec r_j$ is the relative position and $\vec v_{ij}=\vec v_i-\vec v_j$ is the relative velocity. 
In this way, we can estimate the interactions between the intersecting branes from the probe action.
By regarding the probe action for $q_2$-brane in the same background, we estimate the whole interactions between the intersecting branes and we finally obtain the effective action,
\ba
S_{\text{eff}} = \int dt\, (L_\text{kin}+L_\text{1-grav}+L_\text{2-grav}+L_\text{3-grav}+\cdots)
\label{gen-effective-action}
\ea
where
\begin{align}
L_\text{kin} =& \sum_{A=1,2} \sum_{i=1}^{N_A} \left(\frac{m_A}{2}\vect{v}{2}_{A,i}
+\frac{m_A}{8}(\vect{v}{2}_{A,i})^2+\cdots\right)
\\
L_\text{1-grav} =& \sum_{i=1}^{N_1}\sum_{j=1}^{N_2} \frac{\kappa_{d+3}^2 m_1m_2}{d\Omega_{d+1}}\frac{\vect  v2_{ij}}{\vect rd_{ij}}
+\sum_{A=1,2}\sum_{i=1}^{N_A} \frac{\kappa_{d+3}^2m_A^2}{4d\Omega_{d+1}}\frac{\vect  v4_{ij}}{\vect rd_{ij}}
+\sum_{i=1}^{N_1}\sum_{j=1}^{N_2} \frac{\kappa_{d+3}^2m_1 m_2}{2d\Omega_{d+1}}\frac{\vect  v4_{ij}}{\vect rd_{ij}}
+\cdots
\\
L_\text{2-grav} \sim & \sum_{A\neq B}\sum_{i=1}^{N_A}\sum_{j,k=1}^{N_B}\frac{\kappa_{d+1}^4m_Am_B^2}{2d^2\Omega_{d+1}^2}\left(\frac{\vect v4_{ij}}{\vect rd_{ij}\vect rd_{ik}}+\cdots\right)+\cdots
\\
L_\text{3-grav} \sim & \sum_{i,j=1}^{N_1}\sum_{k,l=1}^{N_2}\frac{\kappa_{d+1}^6m_1^2m_2^2}{d^3\Omega_{d+1}^3}\left(\frac{\vect v4_{ij}}{\vect rd_{ij}\vect rd_{ik}\vect rd_{il}}+\cdots\right)+\cdots.
\end{align}
Here $L_\text{n-grav}$ describes the $n$-graviton exchange interactions among $n+1$ branes.
Note that we cannot determine the precise velocity dependence in this method, although it is not a matter for our current purpose.
(In principle we can derive it if we use the method in Ref.~\cite{Okawa:1998pz}.)

By repeating these estimations, we can obtain the interaction terms in the form of infinite series.
However we do not need to derive the whole interactions to discuss the thermodynamics of the intersecting brane system.
As we argue in section \ref{sec:twobrane}, only a few of the interaction terms 
are relevant at low energy ($|\vec{v}_{ij}| \ll 1$).
There $\cQ_1/r^d,~\cQ_2/r^d \gg 1$  would be satisfied as discussed around eq.~(\ref{r-Q}), and, to read off the dominant interactions in this regime from the probe action, we can approximate $H_A \sim \cQ_A/r^d $.
Then we obtain the relevant terms of the effective action
\ba
S_\text{eff} \sim \int dt \, \sum_n 
\sum_{i_1, \dots, i_n}^{N_1}\sum_{j_1, \dots, j_n}^{N_2}
 \frac{\kappa_{{d}+3}^{2(2n-1)} m_1^n m_2^n }{\Omega_{{d}+1}^{2n-1}}
\left(
 \vect{v}{2n} \prod_{k=2}^n \prod_{l=1}^n
 \frac{ 1 }{  \vect{r}{{d}}_{i_1 i_k} \vect{r}{{d}}_{i_1 j_l} }  + \cdots \right).
\ea

For the case of general $M$, the discussion is almost parallel to the $M=2$ case.
We can show that $\cQ_A/r^d \gg 1$ would be satisfied at low energy and we can use the approximation $H_A \sim \cQ_A/r^d $ when we read off the dominant terms of the effective action for the intersecting branes from the probe action (\ref{probe-gen-1}).
We can treat the momentum $P$ in the same way since it can be regarded as an additional charge via the KK reduction.
Especially the gravitational wave which carries the momentum $1/R$ can be regarded as a 0-brane with mass $1/R$ and we take into account the interactions to the other branes.
As a result, we obtain the low energy effective action 
\ba
S_{\text{eff}} = \int dt \,\sum_{n=1}^\infty L_n
\ea
where the first two terms are
\begin{align}
L_1 \sim& 
\sum_{i=1}^{N_1}\sum_{j=1}^{N_2}\cdots \sum_{k=1}^{N_M}
\frac{\kappa_{d+3}^{2(M-1)}\,\hat\prod_A\, m_A}{\Omega_{d+1}^{M-1}}
\left(\frac{\vect v2_{ij}}{\vect rd_{ij}\cdots \vect rd_{ik}}+\cdots\right)
\,,\nt
L_2 \sim& 
\sum_{i_1,i_2=1}^{N_1}\sum_{j_1,j_2=1}^{N_2}\cdots \sum_{k_1,k_2=1}^{N_M}
\frac{\kappa_{d+3}^{2(2M-1)}\,\hat\prod_A\,m_A^2}{\Omega_{d+1}^{2M-1}}
\left(\frac{\vect v4_{i_1j_1}}{\vect rd_{i_1i_2}\vect rd_{i_1j_1}\vect rd_{i_1j_2}\cdots \vect rd_{i_1k_1}\vect rd_{i_1k_2}}+\cdots\right)
\end{align}
and a general $n$-th term is 
\begin{align}
L_n \sim 
\sum_{i_1,\ldots,i_n}^{N_1}\sum_{j_1,\ldots,j_n}^{N_2}\cdots \sum_{k_1,\ldots,k_n}^{N_M}
\frac{\kappa_{d+3}^{2(nM-1)}\,\hat\prod_A\,m_A^n}{\Omega_{d+1}^{nM-1}}
\left(\vect v{2n}\prod_{a=2}^n\prod_{b=1}^n\cdots\prod_{c=1}^n\frac{1}{\vect rd_{i_1i_a}\vect rd_{i_1j_b}\cdots \vect rd_{i_1k_c}}+\cdots\right).
\end{align}
This describes the $nM-1$ graviton exchange among $n$ $q_A$-branes ($A=1,\ldots,M$).

%
\bibliographystyle{utphys}
\bibliography{D1D5}

\providecommand{\href}[2]{#2}\begingroup\raggedright\begin{thebibliography}{10}

\bibitem{Morita:2013wfa}
T.~Morita, S.~Shiba, T.~Wiseman, and B.~Withers, ``{Warm p-soup and near
  extremal black holes},''
  \href{http://dx.doi.org/10.1088/0264-9381/31/8/085001}{{\em
  Class.Quant.Grav.} {\bfseries 31} (2014) 085001},
\href{http://arxiv.org/abs/1311.6540}{{\ttfamily arXiv:1311.6540 [hep-th]}}.

\bibitem{Wiseman:2013cda}
T.~Wiseman, ``{On black hole thermodynamics from super Yang-Mills},''
  \href{http://dx.doi.org/10.1007/JHEP07(2013)101}{{\em JHEP} {\bfseries 1307}
  (2013) 101},
\href{http://arxiv.org/abs/1304.3938}{{\ttfamily arXiv:1304.3938 [hep-th]}}.

\bibitem{Morita:2013wla}
T.~Morita and S.~Shiba, ``{Thermodynamics of black M-branes from SCFTs},''
  \href{http://dx.doi.org/10.1007/JHEP07(2013)100}{{\em JHEP} {\bfseries 1307}
  (2013) 100},
\href{http://arxiv.org/abs/1305.0789}{{\ttfamily arXiv:1305.0789 [hep-th]}}.

\bibitem{Morita:2014ypa}
T.~Morita, S.~Shiba, T.~Wiseman, and B.~Withers, ``{Moduli dynamics as a
  predictive tool for thermal maximally supersymmetric Yang-Mills at large
  N},''
\href{http://arxiv.org/abs/1412.3939}{{\ttfamily arXiv:1412.3939 [hep-th]}}.

\bibitem{Horowitz:1997fr}
G.~T. Horowitz and E.~J. Martinec, ``{Comments on black holes in matrix
  theory},'' \href{http://dx.doi.org/10.1103/PhysRevD.57.4935}{{\em Phys.Rev.}
  {\bfseries D57} (1998) 4935--4941},
\href{http://arxiv.org/abs/hep-th/9710217}{{\ttfamily arXiv:hep-th/9710217
  [hep-th]}}.

\bibitem{Li:1997iz}
M.~Li, ``{Matrix Schwarzschild black holes in large N limit},'' {\em JHEP}
  {\bfseries 9801} (1998) 009,
\href{http://arxiv.org/abs/hep-th/9710226}{{\ttfamily arXiv:hep-th/9710226
  [hep-th]}}.

\bibitem{Banks:1997tn}
T.~Banks, W.~Fischler, I.~R. Klebanov, and L.~Susskind, ``{Schwarzschild black
  holes in matrix theory. 2.},'' {\em JHEP} {\bfseries 9801} (1998) 008,
\href{http://arxiv.org/abs/hep-th/9711005}{{\ttfamily arXiv:hep-th/9711005
  [hep-th]}}.

\bibitem{Li:1998ci}
M.~Li and E.~J. Martinec, ``{Probing matrix black holes},''
\href{http://arxiv.org/abs/hep-th/9801070}{{\ttfamily arXiv:hep-th/9801070
  [hep-th]}}.

\bibitem{Smilga:2008bt}
A.~Smilga, ``{Comments on thermodynamics of supersymmetric matrix models},''
  \href{http://dx.doi.org/10.1016/j.nuclphysb.2009.03.023}{{\em Nucl.Phys.}
  {\bfseries B818} (2009) 101--114},
\href{http://arxiv.org/abs/0812.4753}{{\ttfamily arXiv:0812.4753 [hep-th]}}.

\bibitem{Morita:2014cfa}
T.~Morita and S.~Shiba, ``{Microstates of D1-D5(-P) black holes as interacting
  D-branes},'' \href{http://dx.doi.org/10.1016/j.physletb.2015.05.069}{{\em
  Phys.Lett.} {\bfseries B747} (2015) 164--168},
\href{http://arxiv.org/abs/1410.8319}{{\ttfamily arXiv:1410.8319 [hep-th]}}.

\bibitem{Callan:1996dv}
C.~G. Callan and J.~M. Maldacena, ``{D-brane approach to black hole quantum
  mechanics},'' \href{http://dx.doi.org/10.1016/0550-3213(96)00225-8}{{\em
  Nucl.Phys.} {\bfseries B472} (1996) 591--610},
\href{http://arxiv.org/abs/hep-th/9602043}{{\ttfamily arXiv:hep-th/9602043
  [hep-th]}}.

\bibitem{David:2002wn}
J.~R. David, G.~Mandal, and S.~R. Wadia, ``{Microscopic formulation of black
  holes in string theory},''
  \href{http://dx.doi.org/10.1016/S0370-1573(02)00271-5}{{\em Phys.Rept.}
  {\bfseries 369} (2002) 549--686},
\href{http://arxiv.org/abs/hep-th/0203048}{{\ttfamily arXiv:hep-th/0203048
  [hep-th]}}.

\bibitem{Argurio:1997gt}
R.~Argurio, F.~Englert, and L.~Houart, ``{Intersection rules for p-branes},''
  \href{http://dx.doi.org/10.1016/S0370-2693(97)00205-0}{{\em Phys.Lett.}
  {\bfseries B398} (1997) 61--68},
\href{http://arxiv.org/abs/hep-th/9701042}{{\ttfamily arXiv:hep-th/9701042
  [hep-th]}}.

\bibitem{Khviengia:1997rh}
N.~Khviengia, Z.~Khviengia, H.~Lu, and C.~Pope, ``{Towards a field theory of F
  theory},'' \href{http://dx.doi.org/10.1088/0264-9381/15/4/005}{{\em
  Class.Quant.Grav.} {\bfseries 15} (1998) 759--773},
\href{http://arxiv.org/abs/hep-th/9703012}{{\ttfamily arXiv:hep-th/9703012
  [hep-th]}}.

\bibitem{Martinec:1999sa}
E.~J. Martinec and V.~Sahakian, ``{Black holes and five-brane
  thermodynamics},'' \href{http://dx.doi.org/10.1103/PhysRevD.60.064002}{{\em
  Phys.Rev.} {\bfseries D60} (1999) 064002},
\href{http://arxiv.org/abs/hep-th/9901135}{{\ttfamily arXiv:hep-th/9901135
  [hep-th]}}.

\bibitem{Itzhaki:1998dd}
N.~Itzhaki, J.~M. Maldacena, J.~Sonnenschein, and S.~Yankielowicz,
  ``{Supergravity and the large N limit of theories with sixteen
  supercharges},'' \href{http://dx.doi.org/10.1103/PhysRevD.58.046004}{{\em
  Phys.Rev.} {\bfseries D58} (1998) 046004},
\href{http://arxiv.org/abs/hep-th/9802042}{{\ttfamily arXiv:hep-th/9802042
  [hep-th]}}.

\bibitem{Okawa:1998pz}
Y.~Okawa and T.~Yoneya, ``{Multibody interactions of D particles in
  supergravity and matrix theory},''
  \href{http://dx.doi.org/10.1016/S0550-3213(98)00700-7}{{\em Nucl.Phys.}
  {\bfseries B538} (1999) 67--99},
\href{http://arxiv.org/abs/hep-th/9806108}{{\ttfamily arXiv:hep-th/9806108
  [hep-th]}}.

\end{thebibliography}\endgroup

%

\end{document}